\documentclass[preprint,showkeys,preprintnumbers]{revtex4}

\newcommand{\pt}{\ensuremath{p_T}}
\newcommand{\zpt}{\ensuremath{\pt^Z}}

\newcommand{\DYJets}{{\it DYJets}}
\newcommand{\DoubleMu}{{\it DoubleMu}}
\newcommand{\SumPt}{\ensuremath{\sum \pt}}
\newcommand{\Nch}{\ensuremath{N_{ch}}}

\usepackage{graphicx}% Include figure files
\usepackage{dcolumn}% Align table columns on decimal point
\usepackage{bm}% bold math
\usepackage{epsfig}
\usepackage{epstopdf}
\usepackage{grffile}
\usepackage{color}
\usepackage{colordvi}
\usepackage{amssymb}
\usepackage{rotating}
\usepackage{lscape}
\usepackage{float}

\usepackage{amsmath}
\usepackage{subfigure} 

\usepackage{footnote}
 \makesavenoteenv{environmentname}

\usepackage{hyperref}

\usepackage[T1]{fontenc} % if needed

\begin{document}

%\linenumbers

\title{Explicit Jet Veto as a Tool to Purify the Underlying Event in the Drell-Yan Process Using CMS Open Data}

\author{Saeid Paktinat Mehdiabadi}
\email{spaktinat@yazd.ac.ir}
\affiliation{Faculty of Physics, Yazd University, P.O. Box 89195-741, Yazd, Iran\\
	School of Particles and Accelerators, Institute for Research in Fundamental Sciences (IPM), P.O.Box 19395-5531, Tehran, Iran}

\author{Ali Fahim}
\email{a.fahim@ut.ac.ir}
\affiliation{Department of Engineering Science, College of Engineering, University of Tehran, P.O. Box 11155-4563, Tehran, Iran}

\date{\today}

\begin{abstract}
The underlying event is an important part of high-energy collision events. In the event generators, the underlying event is tuned by fits to collision data. Usually, the underlying event observables are affected by the existence of extra jets and it is difficult to find a part of the phase space which is dominated by the underlying event. In this paper, we suggest to veto the jets in the considered region to disentangle these effects. The idea is verified to work on CMS Open Data. To our knowledge, it is the first time that such ideas are tested on real collision data.
\end{abstract}

\maketitle
\section{Introduction}\label{sec:int} 
In the hadron colliders, the physics process of our interest is usually a hard scattering of a single parton from another one. 
However, in each proton-proton (pp) collision, in addition to such primitive hard-scattering products, there are many particles originating either from the soft color interactions between the rest of the partons, or from additional hard scattering of other partons at the same pp collision \cite{Sjostrand:1986ep}. While the former parts are called beam remnants, the latter ones are known as the multiple partonic interactions (MPI). Both processes are called the underlying event (UE). 
Another intrinsic part of the hadron collisions, is the gluon radiation occurring either before the original hard scattering (initial state radiation (ISR)) or after that (final state radiation (FSR)).  
Particles coming from ISR and FSR mimic the UE activities. Hence, it is difficult to empirically distinguish UE from the ISR and FSR productions.

A precise measurement or observation of a new physics will not be possible without a good understanding of these processes in a hadron collider. Due to low transverse momentum (\pt) of their products, perturbative QCD may not accurately describe these processes. So, the practical understanding of UE is limited to the phenomenological models. 
The CDF \cite{Affolder:2001xt,Field:2002vt,Acosta:2004wqa,Aaltonen:2010rm} experiment at Tevatron and 
the ATLAS \cite{Aad:2010fh,Aad:2011qe,Aad:2012jba,Aad:2014hia,Aad:2014jgf,Aaboud:2017fwp}, ALICE \cite{ALICE:2011ac} and CMS \cite{Khachatryan:2010pv,Chatrchyan:2011id,Chatrchyan:2012tb,Chatrchyan:2013ala,Sirunyan:2017vio} experiments at LHC have carried out several measurements of parameters and observables sensitive to UE at their available center-of-mass energies ($\sqrt{s}$).
Due to easy identification of Z-boson production, the experiments usually study the UE activities in the events containing the Drell-Yan(DY) process. 

\begin{figure}[!htb]
	\centering
	\includegraphics[width=.49\textwidth]{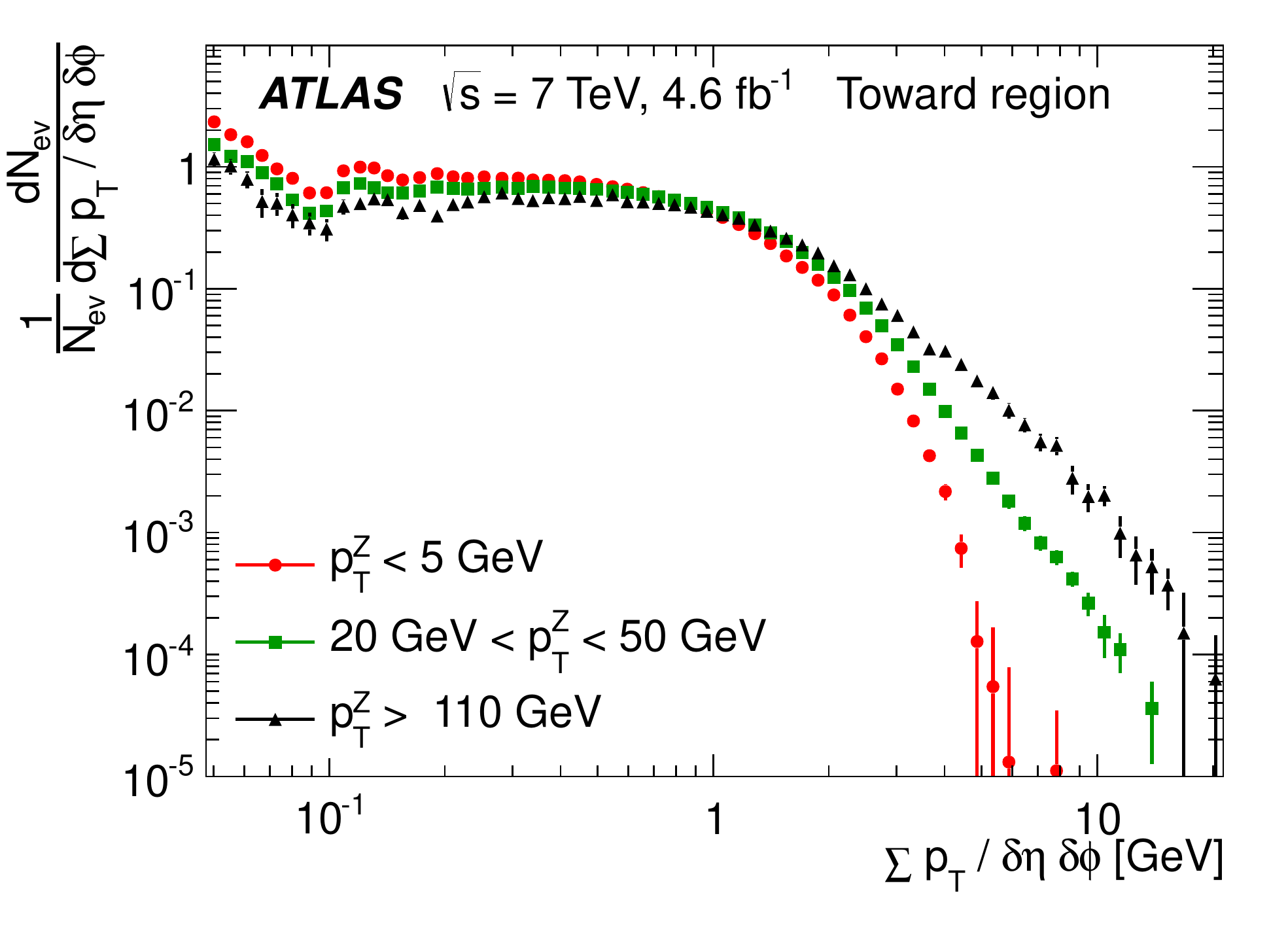}
	\includegraphics[width=.49\textwidth]{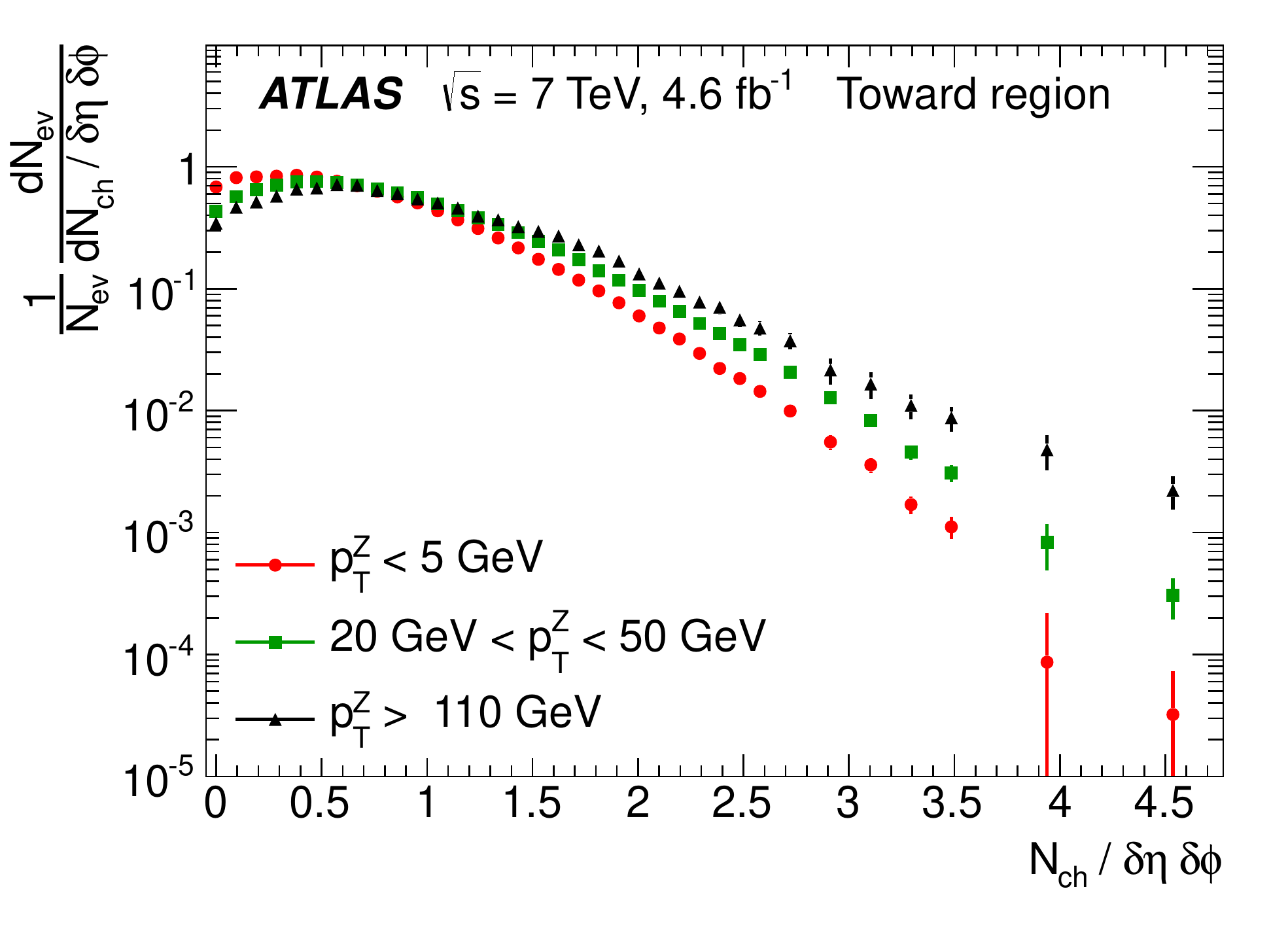}
	\caption{Distributions of the scalar transverse momentum sum density and charged particle multiplicity density. The figure illustrates the distributions in three different intervals of \zpt\cite{Aad:2014jgf}.}\label{fig_atlas_05a_09a}
\end{figure}

To find UE  parameters, it is important to have a clean environment dominated by the UE activities. Usually, it is difficult to find this environment and 
extra jets affect the UE observables.
For example, two of them which are investigated by the ATLAS experiment \cite{Aad:2014jgf} are shown in Fig. \ref{fig_atlas_05a_09a}.
In a pure UE environment, the shown distributions must be the same for different  Z-boson \pt(\zpt), but obviously it is not the case here.
Recently, it is suggested to use the event shape observables to find the UE dominated regions \cite{Kar:2018uno}.
Here, we propose a new method as an alternative procedure to specify a cleaner UE environment. In the new method, the jets are vetoed explicitly from the region of interest.
For the first time, to verify the performance of the methods, Open-Data \cite{Mccauley:2016wfz} collected or simulated by the CMS experiment is used. The events contain a muon pair consistent with a Z-boson production in pp collisions at $\sqrt{\textrm{s}}$ = 7 TeV.

After the introduction, the paper is structured as follows.
Section \ref{sec:opendata} describes the simulated and collision data used in this analysis. 
The method and the results are presented in section \ref{sec:results}. Then section \ref{sec:con} concludes the paper. 

\section{CMS Open Data}\label{sec:opendata} 
The LHC experiments have released part of their pp collision data and the related Monte Carlo simulations as a project known as the Open Data. 
In this analysis, we look at the events with two opposite charge muons in $\sqrt{s}$ = 7 TeV taken by the CMS experiment in 2011. The datasets for collision data and simulated events are shown in Tab. \ref{tab:Datasets}.
\begin{table}[!htb]
	\centering
	\caption{The path of simulated and collision data in the CMS Open Data storage. The abbreviation and number of events analyzed  in the paper are also shown.}
	\begin{tabular}{|c|c|c|}
		\hline
		Dataset Path & Abbreviation & Number of Events \\\hline
		/MonteCarlo2011/Summer11LegDR... &  &\\
		/DYJetsToLL\_TuneZ2\_M-50\_7TeV-madgraph-tauola... &  &\\
		/AODSIM/PU\_S13\_START53\_LV6-v1/XXXX & \DYJets &10169936\\
		XXXX $\in$ \{00000, 010000, 010001, 010002\} &  &\\\hline
		/Run2011A/DoubleMu/AOD/12Oct2013-v1/10000/ & \DoubleMu  &26320428\\\hline
	\end{tabular}
	\label{tab:Datasets}
\end{table}
Due to limitation of the resources, only part of the available data is used in the analysis. To read the data files, the official software of CMS (CMSSW\_5\_3\_32) is used as explained in the guidelines provided by the experiment\footnote{http://opendata.cern.ch}.

Every event is required to have at least one vertex, which is built from a fit to extrapolation of tracks to the nominal collision point. The  fitted vertex must have more than 4 degrees of freedom. The distance of the vertex from the nominal collision point can be at most, 24 cm in z-direction and 2 cm in $\rho$-direction, where the cylinderical coordinate has an origin at the center of the detector and z-direction is parallel to the beams.

The objects used in the analysis are defined as follows:
\begin{itemize}
	\item{Muons:} The muon candidates must pass \verb|isPFMuon| and either \verb|isGlobalMuon| or  \verb|isTrackerMuon|. 
	\item{Electrons:} The electron candidates are required to pass a tight identification 
	(\verb|electronID("eidRobustTight")|).
	\item{Jets:} The jets are reconstructed using the anti-$k_T$ algorithm with a distance parameter 0.5. To clean the jets, the following cuts are applied: \verb|numberOfDaughters| $>$ 1, 
	\verb|neutralHadronEnergyFraction| $<$ 0.99 and \verb|neutralEmEnergyFraction| $<$ 0.99. For jets in the tracker coverage (pseudorapidity |$\eta$| $<$ 2.4) these extra conditions are also required: \verb|chargedEmEnergyFraction| $<$ 0.99, \verb|chargedHadronEnergyFraction| $>$ 0 and \verb|chargedMultiplicity| $>$ 0.
	\item{Tracks:} Among all tracks, only the following tracks are selected. The measured values of $\rho$ and z with respect to the vertex of the hardest muon over their uncertainty must be less than 3 and the uncertainty of \pt~over its value must be less  than 0.05. The former condition guarantees that only the tracks from the main vertex enter the analysis and the tracks from other soft collisions (pile up tracks) are rejected. The tracks have to pass also a \verb|HighPurity| identification quality.
\end{itemize}
The selected leptons (electrons or muons) need also to have $\pt > 25$ GeV and $|\eta|<$ 2.5. If the sum of energy deposit close to a lepton in the tracker (\verb|trackIso|) and calorimeters (\verb|caloIso|) is less than 20\% of lepton \pt, it enters the analysis as an isolated lepton.
The selected jets pass the kinematic cuts of $\pt >$ 20 GeV and |$\eta$| $<$ 2.5. The selected tracks have the same cut on |$\eta$| with $\pt >$ 0.5 GeV.

For both \DYJets~and \DoubleMu~datasets, the following selections are applied. 
\begin{itemize}
	\item{DiMuons:} To select the Z boson candidates decaying to dimuon pair, every event is required to have at least one pair of opposite charge muons with an invariant mass close to the Z boson mass (65-115 GeV).
	\item{Electron veto:} To reject the backgrounds from diboson production, there should not be any isolated electron in the event. It is also useful to further reduce the size of the datasets.
	\item{Jet:} To have boosted Z bosons, at least one jet must be in the event.
\end{itemize}

For \DoubleMu~sample, only parts of data, certified by the experiment, are used.
For \DYJets~sample, both generated charged particles and reconstructed tracks are used in the analysis. The former part can provide the generated information without the complexities introduced by the non perfect detectors.

\section{Analysis and Results}\label{sec:results}
\begin{figure}[!htb]
	\centering
	\includegraphics[width=.35\textwidth]{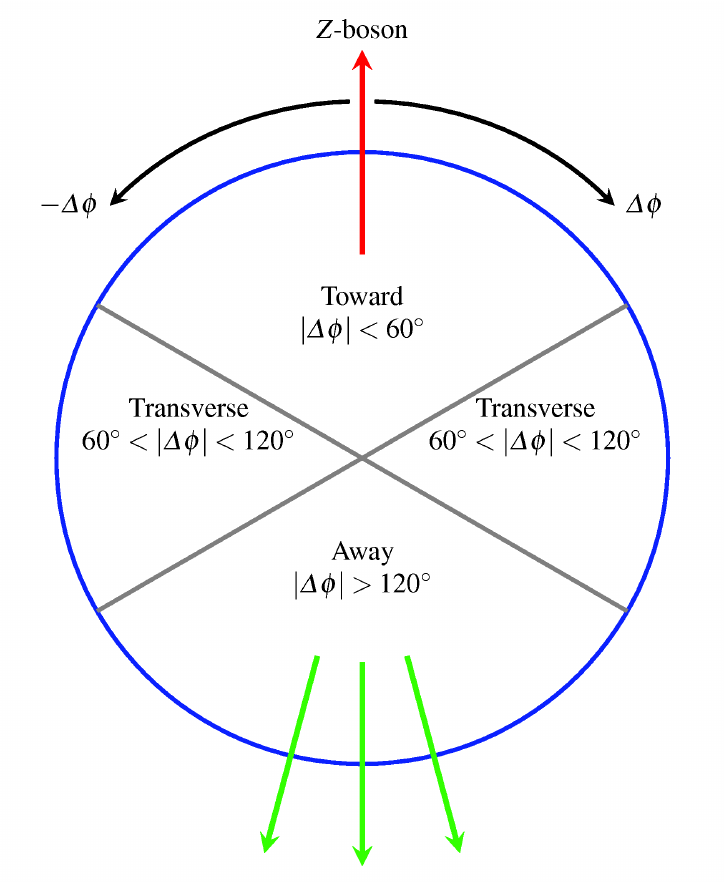}
	\caption{Definition of UE regions in the azimuthal plane regarding the Z-boson direction \cite{Aad:2014jgf}.}\label{fig_regions}
\end{figure}

The different regions of the azimuthal plane respect to the Z-boson flight direction could have different sensitivity to the UE activities. 
For the comparability, the same definition of the UE regions as in the ATLAS paper \cite{Aad:2014jgf} (Fig. \ref{fig_regions}) is used.

The charged tracks, depending on their azimuthal distance with respect to the Z-boson direction ($|\Delta\phi|$), 
are divided in the following three azimuthal UE regions:
\begin{itemize}%\setlength\itemsep{-0.5em}
	\item $|\Delta\phi|<60^{\circ}$, the {\bf  Toward region}, 
	\item $60^{\circ}<|\Delta\phi|<120^{\circ}$, the {\bf  Transverse region}, and 
	\item $|\Delta\phi|>120^{\circ}$, the {\bf  Away region}.
\end{itemize}

\begin{figure}[!htb]
	\centering
	\includegraphics[width=.49\textwidth]{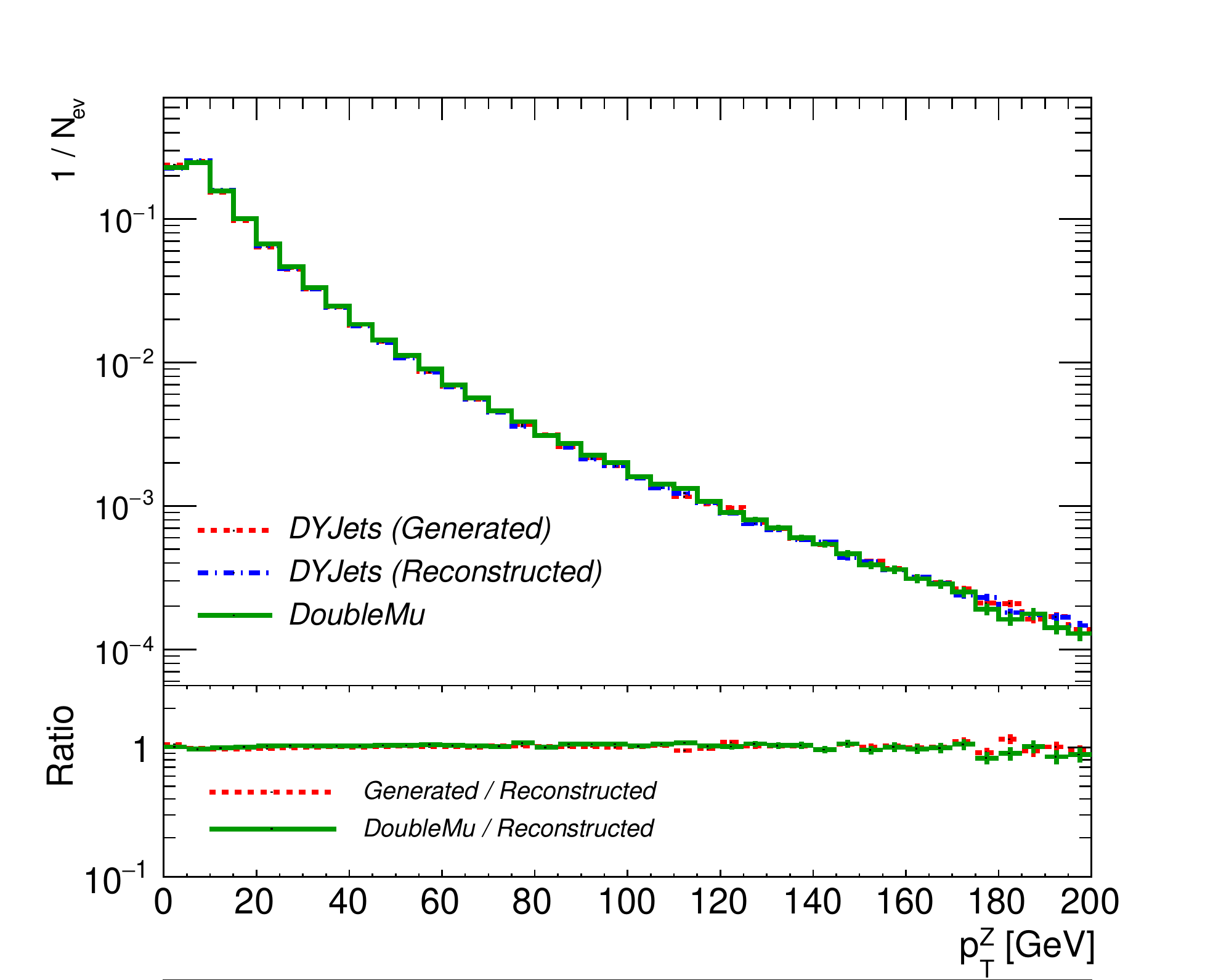}
	\includegraphics[width=.49\textwidth]{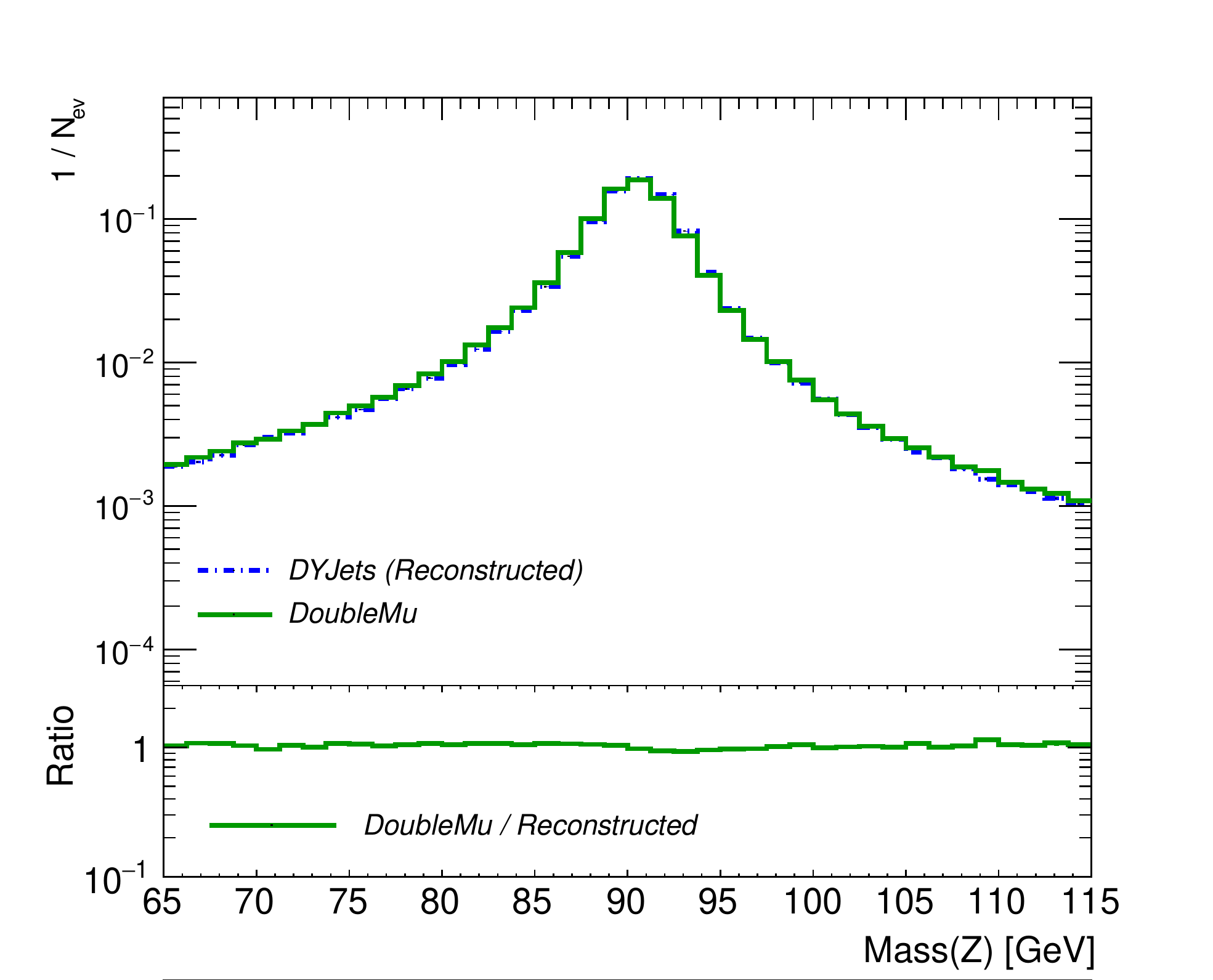}
	\caption{Distribution of \zpt~for simulated samples in generator and reconstruction level and the data collected by the CMS experiment. Most of the Z-bosons have a low \pt. In the right side, the Z lineshape is compared between reconstructed events from \DYJets~and \DoubleMu~samples.}\label{fig_zpt}
\end{figure}

A considerable momentum of Z-boson could make this separation between the regions a lot more clear.
However, such cases have only a small contribution to entire sample. Figure \ref{fig_zpt} (left)
shows the distribution of \zpt~for the generated and reconstructed events from \DYJets~sample and also the reconstructed events from \DoubleMu~sample.
As the figure illustrates, for example the events with $\zpt \geq 50$ GeV, make only few percent of the selected events. In Fig. \ref{fig_zpt} (right), the invariant mass of the selected dimuon is compared in reconstructed events from \DYJets~and \DoubleMu~samples. There is a good agreement between the distributions, showing a good description of the real data in the Monte Carlo simulated events.

In the analysis, the tracks corresponding to the Z-boson products are omitted. 
Since the hard scattering particles, which balance the transverse momentum of the Z-boson, are mostly in the Away region, all remaining tracks in the toward region are dominantly from the UE activities. 
The charged track multiplicity (\Nch) and the scalar transverse momentum sum  of the charged tracks (\SumPt) are the main observables which are studied below. 
It is common to look at the density of these observables, $\Nch/\delta\eta\delta\phi$ and $\SumPt/\delta\eta\delta\phi$, so \Nch~and \SumPt~are divided by the angular area $\delta\eta\delta\phi$. In this analysis, the observables are investigated only in the toward region, where the angular area is $\delta\eta\delta\phi=10\pi/3$.

\begin{figure}[!htb]
	\includegraphics[width=0.49\textwidth]{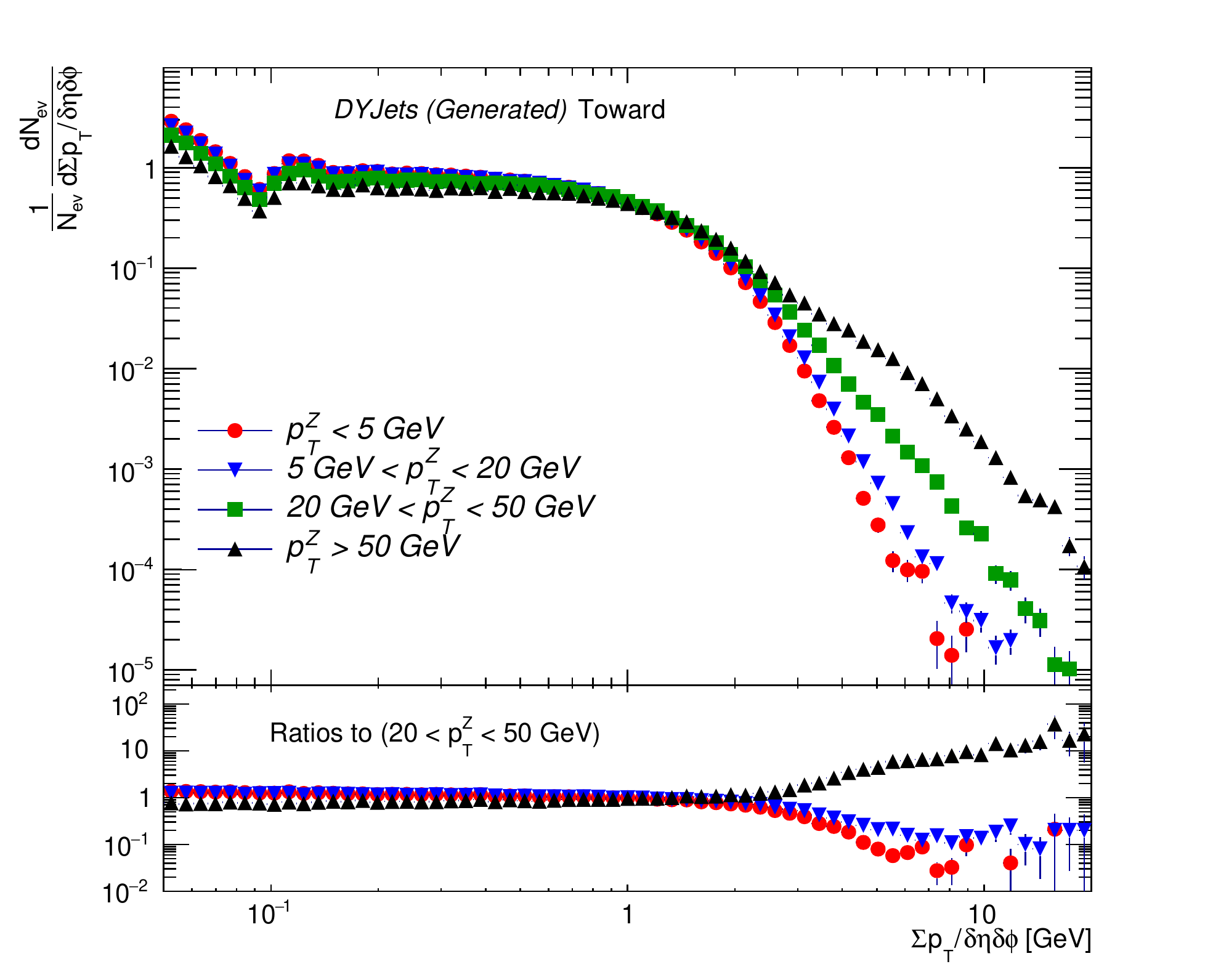}
	\includegraphics[width=0.49\textwidth]{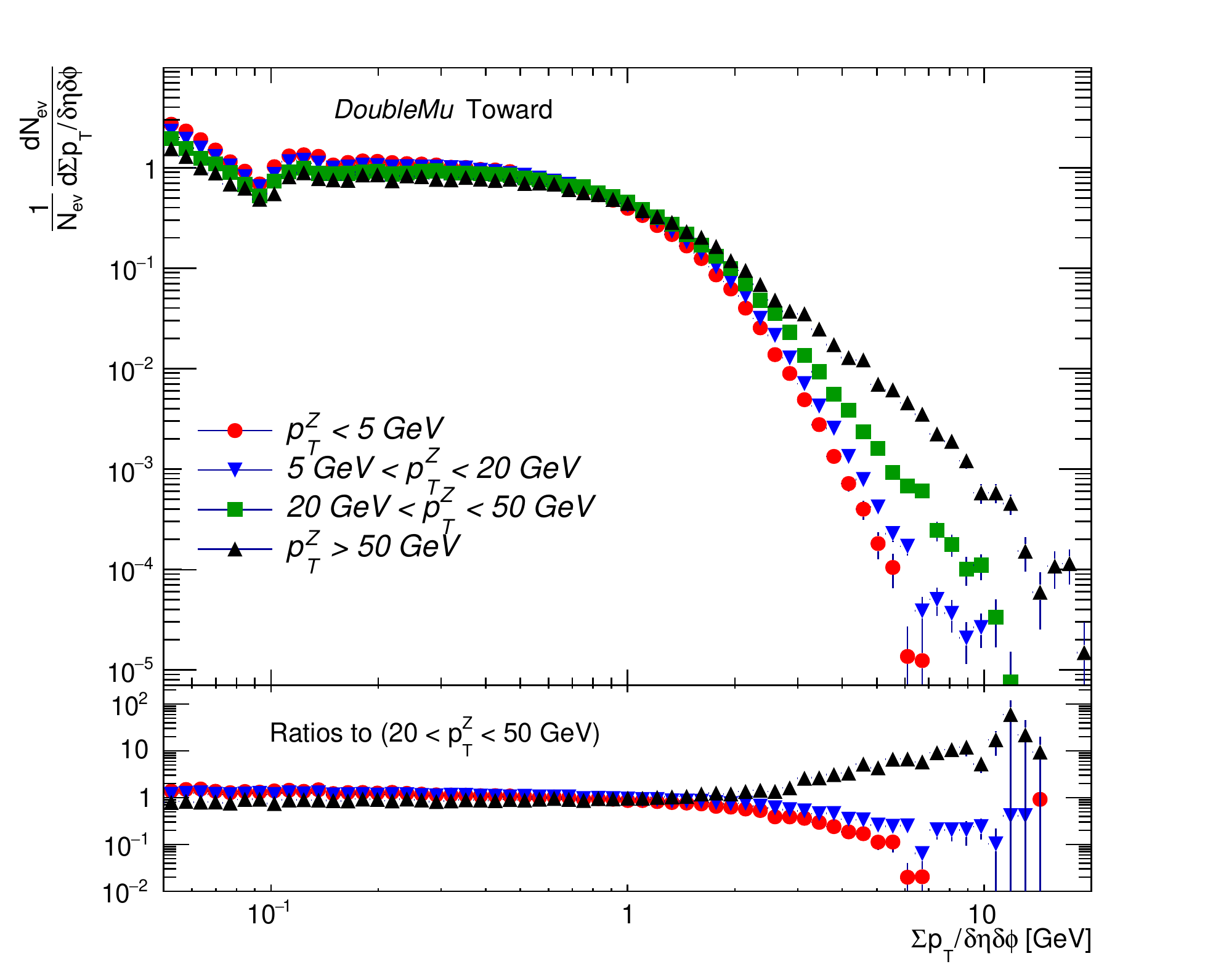}
	\caption{Distributions of \SumPt~density of tracks, in four different \zpt~ranges in the toward region. Distributions are in the generator level (left) and real data (right).}\label{fig_sumpt_dygen_dmrec_nc}
\end{figure}

\begin{figure}[!htb]
	\includegraphics[width=0.49\textwidth]{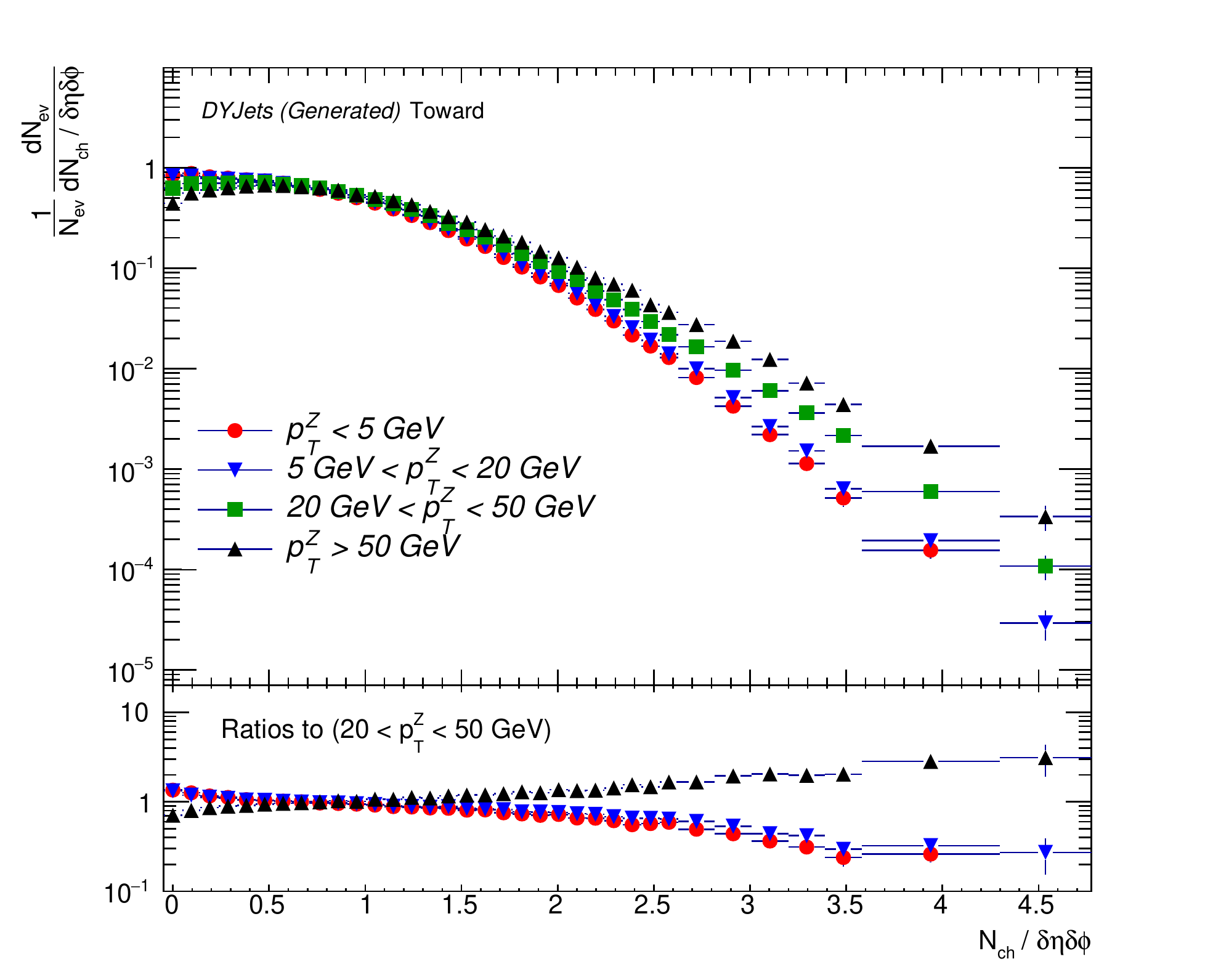}
	\includegraphics[width=0.49\textwidth]{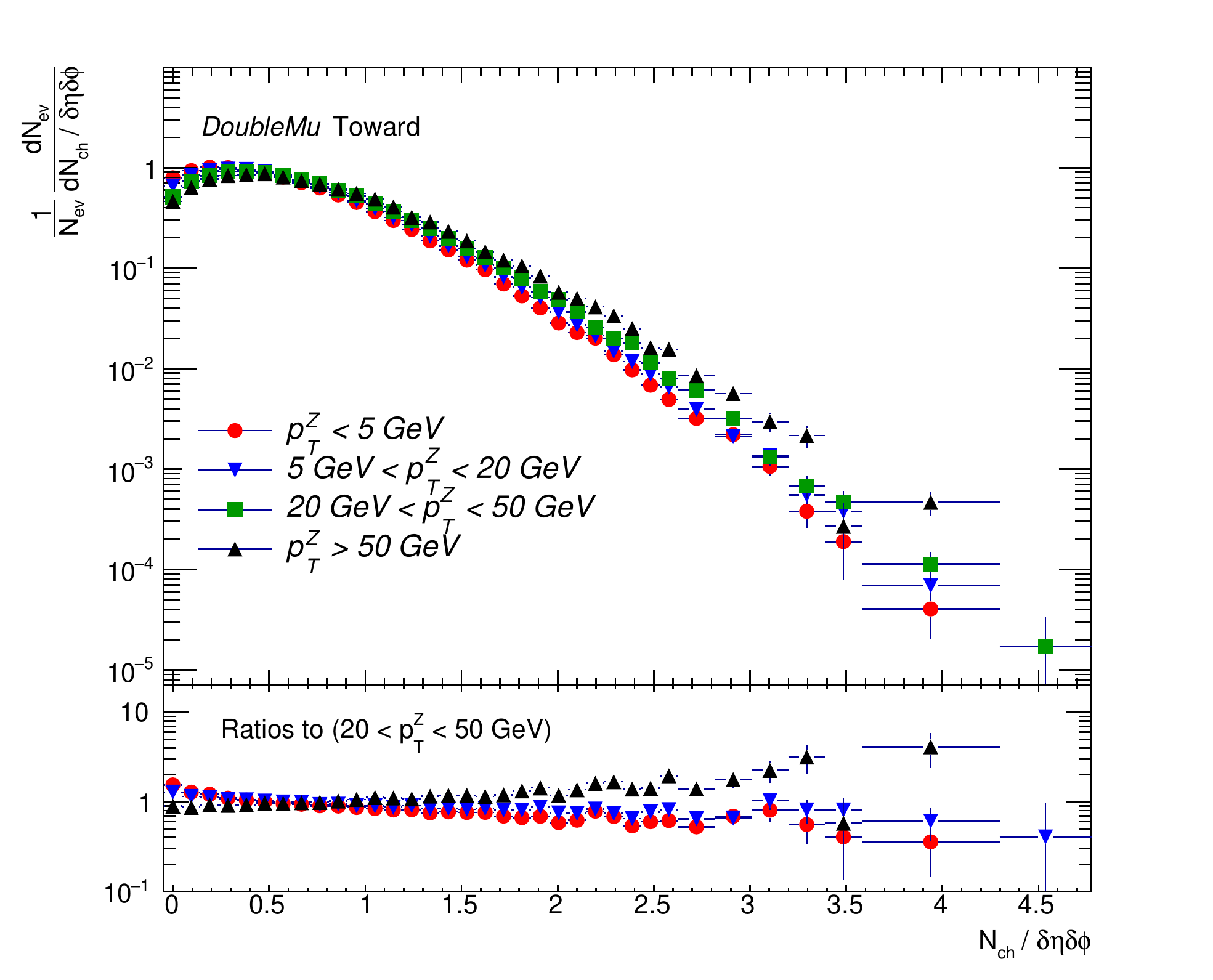}
	\caption{Distributions of \Nch~density, in four different \zpt~ranges in the toward region. Distributions are in the  generator level (left) and real data (right).}\label{fig_nchg_dygen_dmrec_nc}
\end{figure}

Figures \ref{fig_sumpt_dygen_dmrec_nc} and \ref{fig_nchg_dygen_dmrec_nc}
illustrate the distribution of \SumPt~and \Nch~densities, respectively for the toward region. Distributions are provided in four different \zpt~ranges, in both the generator level (left) and real collision data (right).
The figures depict that in contrast to what is expected for a UE dominated region,  distribution of observables are not independent of \zpt~ranges. Following the conclusion of Ref. \cite{Aad:2014jgf}, it seems still the toward region is contaminated by the tracks from additional jets. The distributions for the reconstructed events in \DYJets~sample also show the similar features.

\subsection{Jet Veto}
In Ref.\cite{Kar:2018uno}, the event shapes are suggested as a tool to select a part of the phase space which is enriched by UE. It is not confirmed by our investigation on the available Open Data.
To disentangle the effect of UE from the extra jets, we suggest to veto the jets in the toward region explicitly. The UE sensitive distributions are plotted again after this veto.

\begin{figure}[!htb]
	\centering
	\includegraphics[width=.49\textwidth]{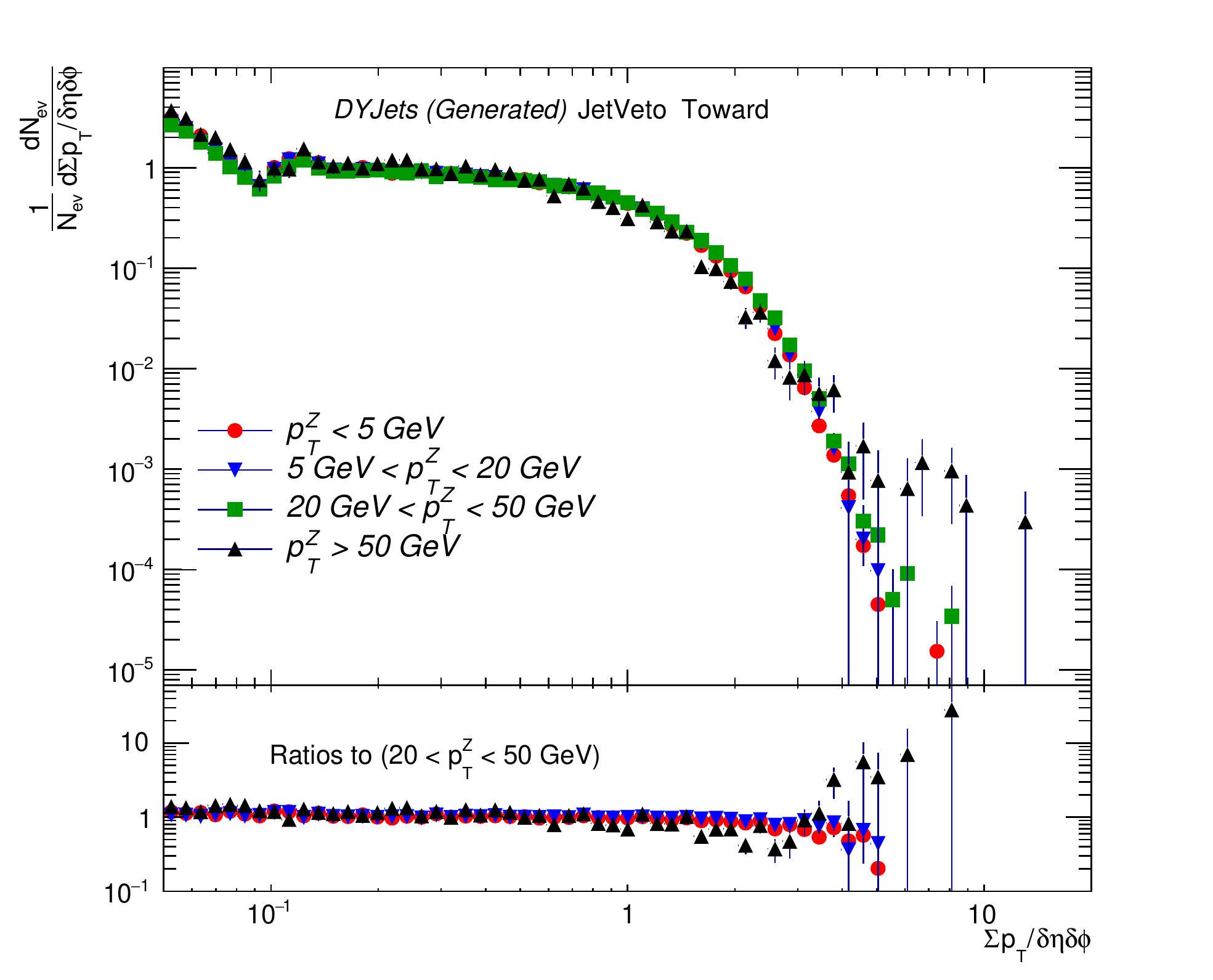}
	\includegraphics[width=.49\textwidth]{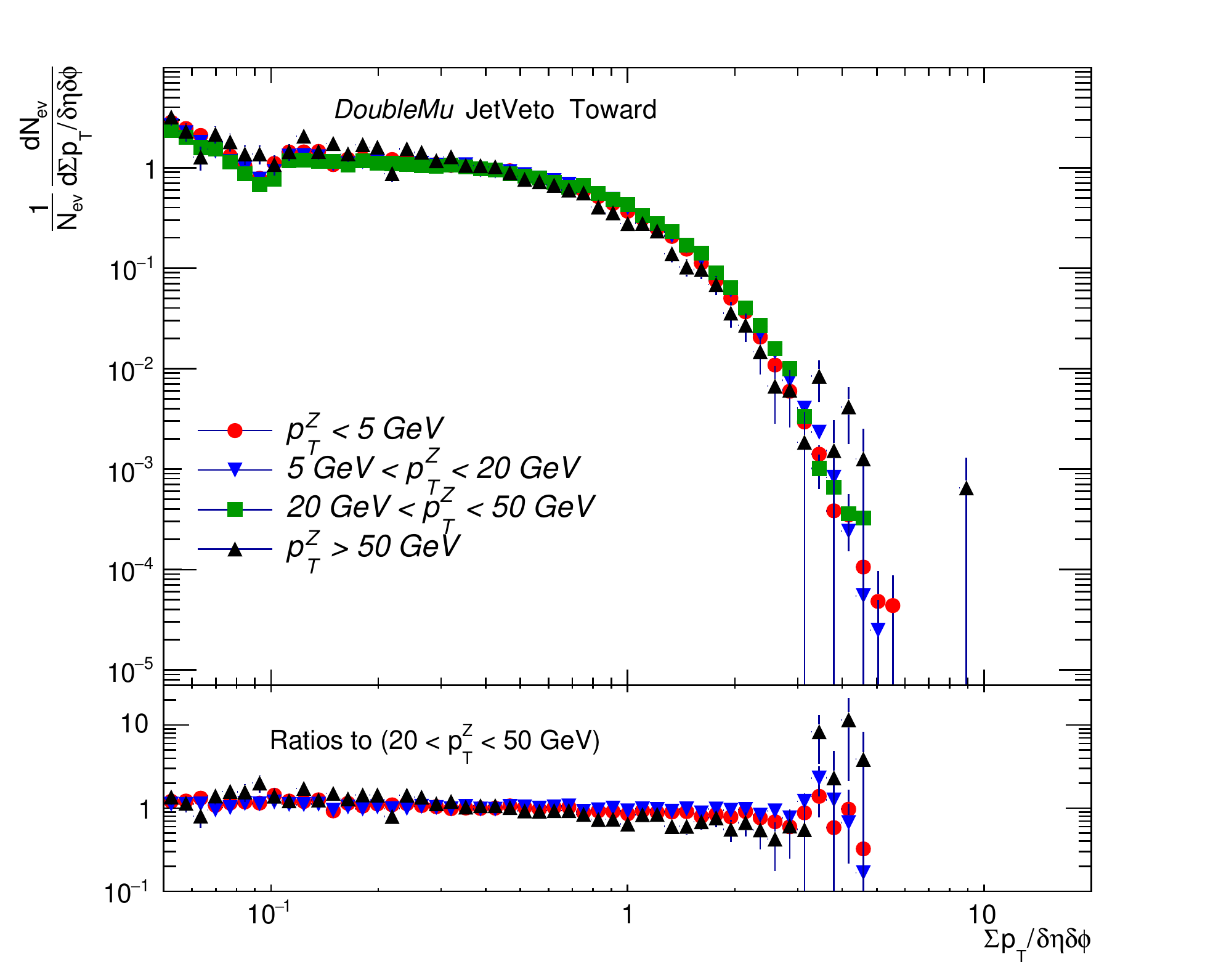}
	\caption{The \SumPt~density in generated events from \DYJets~sample and reconstructed events from \DoubleMu~sample, when jet veto is applied. The expected UE behaviour is observed.}\label{fig_sumpt_dygen_njt}
\end{figure}

The \SumPt~density is shown in Fig. \ref{fig_sumpt_dygen_njt} (left)
for the generated events in \DYJets~sample. The figure should be compared to Figs. \ref{fig_atlas_05a_09a}(left) and \ref{fig_sumpt_dygen_dmrec_nc}.
Apart from the events with high \zpt, which have a low statistics and fluctuations are dominant (See Fig. \ref{fig_zpt}(left)),  it is clear that changing the \zpt~range does not affect the UE distribution as it is expected for a pure UE region. It is concluded that, this region is a UE dominated region and can be used safely to tune the Monte Carlo generators to reproduce the UE observables as correctly as possible.
The same distribution is investigated by the reconstructed events from \DYJets~and \DoubleMu~samples  %\ref{fig_sumpt_dmrec_njt}. 
and the same result is concluded. The latter distribution is shown in Fig. \ref{fig_sumpt_dygen_njt} (right).

\begin{figure}[!htb]
	\centering
	\includegraphics[width=.49\textwidth]{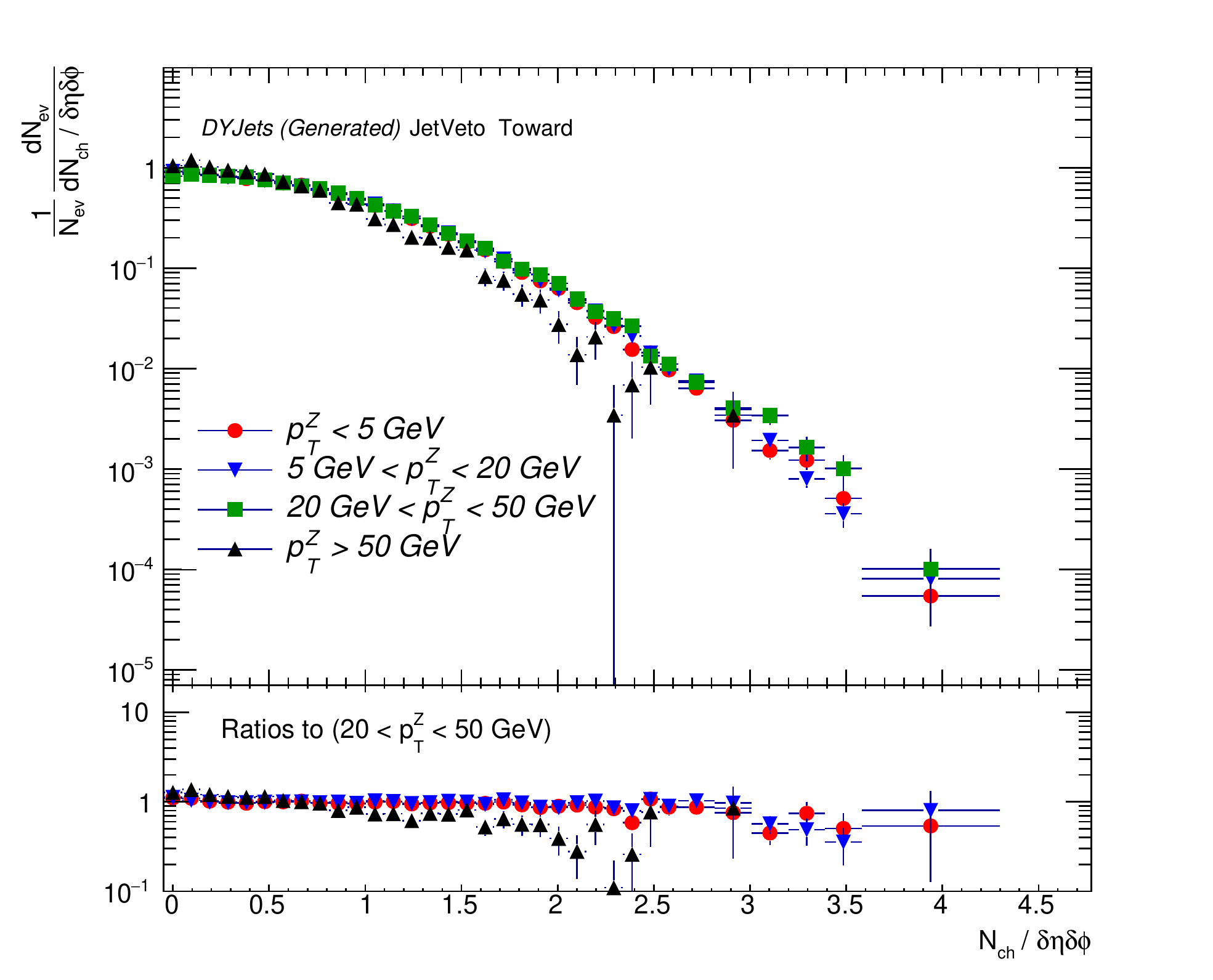}
	\includegraphics[width=.49\textwidth]{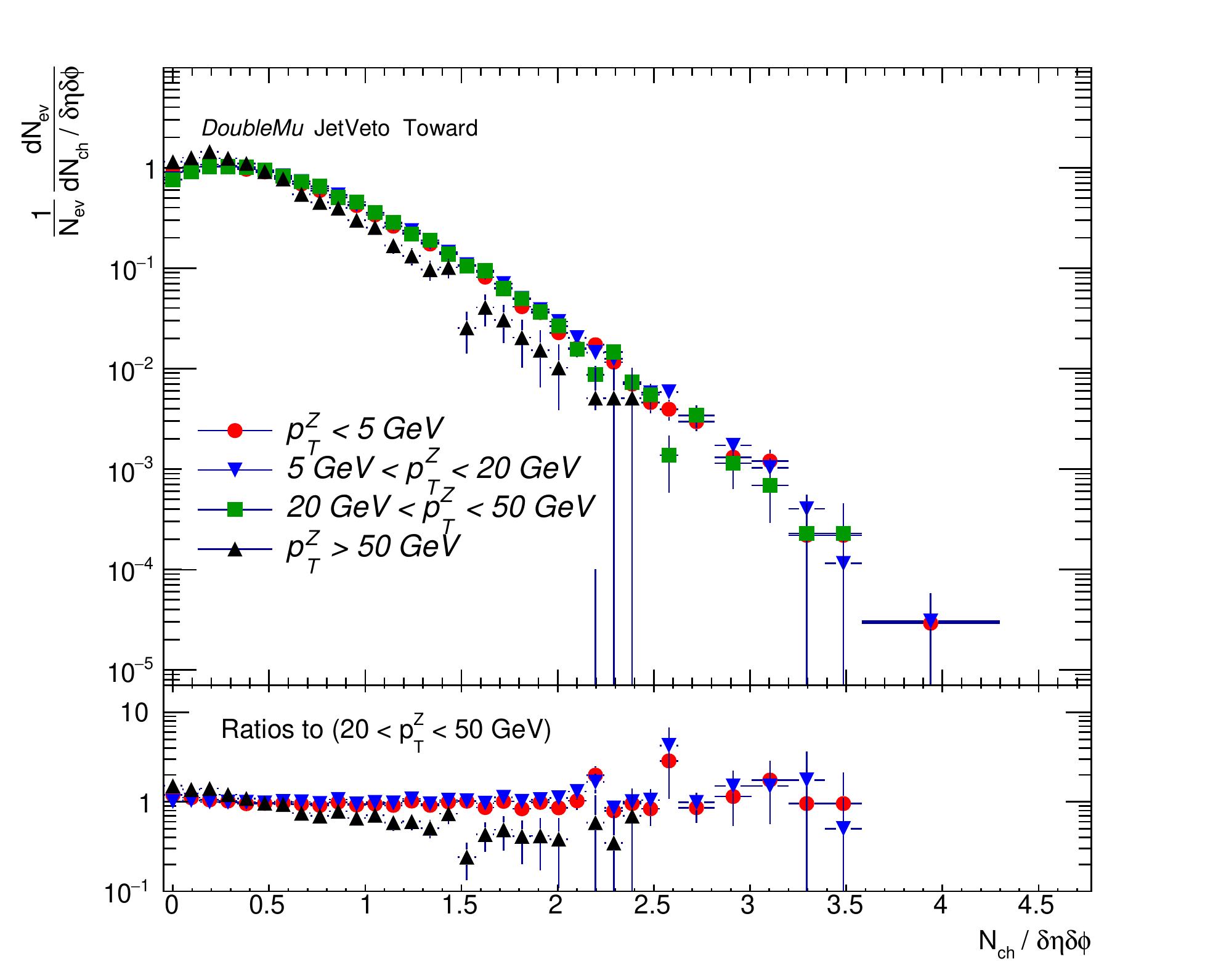}
	\caption{The \Nch~density in generated events of \DYJets~sample and reconstructed events of \DoubleMu~sample, when jet veto is applied. The expected UE behaviour is observed.}\label{fig_nchg_dygen_njt}
\end{figure}

The \Nch~density is another UE sensitive observable which is studied in this analysis. Vetoing the jets from the toward region can suppress the effect of the extra jets in \Nch~density distribution also. The distribution for generated events in \DYJets~sample  is shown in Fig. \ref{fig_nchg_dygen_njt} (left).
Comparing to Figs.  \ref{fig_atlas_05a_09a}(right) and \ref{fig_nchg_dygen_dmrec_nc}, the expected behaviour for a UE dominated region is seen. 
The same features are seen for the distributions made by the reconstructed events from \DYJets~and 
\DoubleMu~samples. The latter distribution is shown in Fig. \ref{fig_nchg_dygen_njt} (right). 
The different behaviour for high \pt~sample can be assigned to statistical fluctuations. The jet veto is not perfect, because the jet selection is not perfect. The jet identification and kinematic cuts, can introduce some bias in jet veto, but similar distributions for three \zpt~bins, below 50 GeV, can guarantee that the effect for high \zpt~bin is not due to the jet selection. It seems the low \pt~jets and misidentified or nonidentified jets, can not be the responsible of this effect.

It is suggested that the LHC experiments try this proposal and compare different MC tunes for UE sensitive observables. It should be emphasized that, since no comparison is done between reconstructed and generated events, unfolding is not necessary and it is not done in this analysis.

\section{Conclusion}\label{sec:con} 
The underlying event as an important part of high-energy collision events, is tuned in the event generators by fits to collision data. 
It is important to find a part of the phase space which is dominated by UE. 
Different methods are proposed to find such environments. Usually, the UE observables are affected by the existence of extra jets. A new method  is proposed to disentangle the effects of extra jets from UE. It is suggested to veto the extra jets in the toward region. For the first time, the CMS Open Data is used to investigate the performance of this method and an alternative method. After applying the jet veto, the UE sensitive observables show the expected behavior for a fully dominated UE environment. The result is verified in Monte Carlo simulated events (both generated and reconstructed objects) and also real collision data.

\section{Acknowledgments}
The authors are grateful to the CMS and ATLAS collaborations for their fantastic results and releasing the high quality LHC data.

\end{document}